# KURTOSIS OF LARGE–SCALE COSMIC FIELDS

E. L. Łokas [1], R. Juszkiewicz [2], D.H. Weinberg [3] and F.R. Bouchet [4]

## ABSTRACT

An attractive and simple hypothesis for the formation of large–scale structure is that it developed by gravitational instability from primordial fluctuations with an initially Gaussian probability distribution. Non–linear gravitational evolution drives the distribution away from the Gaussian form, generating measurable skewness and kurtosis even when the variance of the fluctuations is much smaller than unity. We use perturbation theory to compute the kurtosis of the mass density field and the velocity divergence field that arises during the weakly non–linear evolution of initially Gaussian fluctuations. We adopt an Einstein–de Sitter universe for the perturbative calculations, and we discuss the generalization to a universe of arbitrary $\Omega$. We obtain semi–analytic results for the case of scale–free, power–law spectra of the initial fluctuations and final smoothing of cosmic fields with a Gaussian filter. We also give an exact analytical formula for the dependence of the skewness of these fields on the power spectrum index. We show that the kurtosis decreases with the power spectrum index, and we compare our more accurate results for the kurtosis to previous estimates from Monte Carlo integrations. We also compare our results to values obtained from cosmological N–body simulations with power–law initial spectra. Measurements of the skewness and kurtosis parameters can be used to test the hypothesis that structure in the universe formed by gravitational instability from Gaussian initial conditions.

[1] Copernicus Astronomical Center, Bartycka 18, 00-716 Warsaw, Poland; e–mail: lokas@camk.edu.pl
[2] Copernicus Astronomical Center, Bartycka 18, 00-716 Warsaw, Poland; e–mail: roman@camk.edu.pl
[3] Institute for Advanced Study, Olden Lane, Princeton, New Jersey 08540, USA; e–mail: dhw@guinness.ias.edu
[4] Institut d'Astrophysique de Paris, 98 bis Bd Arago, 75014 Paris, France; e–mail: bouchet@iap.fr

# 1 Introduction

It is widely believed that the large–scale structure that we observe today evolved by gravitational instability from small–amplitude, primordial fluctuations, generated by some physical process in the very early universe. The most common assumption is that these primordial fluctuations had a Gaussian probability distribution. In addition to its mathematical simplicity, this assumption has a broad physical motivation: the central limit theorem implies that fluctuations emerging from a sum of many uncorrelated, random effects will have a nearly Gaussian distribution. Furthermore, Gaussian fluctuations are predicted by the simplest versions of inflationary cosmology. Convincing evidence *against* Gaussian initial conditions would be extremely important, since a departure from Gaussian form would rule out many scenarios, and the nature of this departure might point us towards a fundamental physical theory for the origin of primordial fluctuations (or, perhaps, to a non–gravitational theory for the origin of structure).

Microwave background anisotropies probe cosmic fluctuations at a time when their amplitude is small, so that their statistical distribution should be close to its primeval form. However, the limited signal–to–noise of existing microwave background measurements, and the insuperable problem of "cosmic variance" in the anisotropies on large angular scales, make it difficult to test the Gaussian hypothesis to high precision with such measurements. An alternative is to use the density and velocity fields of the present–day galaxy distribution, but in this case one must take account of the effects of non–linear gravitational evolution. In particular, the skewness, kurtosis, and higher–order reduced (or connected) moments vanish for a Gaussian distribution, but non–linear evolution distorts the distribution of the density contrast and the velocity divergence, generating non–zero values for these moments. This paper follows the path pioneered by Peebles (1980) and Fry (1984), using perturbation theory to compute the values of these moments that arise during weakly non–linear evolution.

In order to make analytic predictions comparable to observations or to numerical simulations, one needs to introduce smoothing of the fields, which results in dependence of the reduced moments on the index of the initial power spectrum. The third reduced moment – the skewness – was studied by Juszkiewicz et al. (1993a; hereafter JBC) for both the Gaussian and top-hat filters. The fourth reduced moment – the kurtosis – was calculated for the top-hat filter by Bernardeau (1993). For a Gaussian filter, estimates of the kurtosis have been computed by Monte Carlo integration methods, for a cold dark matter spectrum (Goroff et al. 1986), and for power–law spectra (Catelan and Moscardini 1994).

In this paper we present a more exact perturbative calculation of the kurtosis of cosmological density and velocity fields, and we compare our results to those obtained from cosmological N–body simulations. In section 2 we present the perturbative solutions of the dynamical equations of motion for the density contrast and velocity divergence fields of the orders needed. Section 3 is devoted to statistical properties of the density and velocity divergence fields, and to the calculation of the skewness and kurtosis parameters. Section 4 presents results from a series of N–body simulations with Gaussian initial conditions and power–law initial spectra. A concluding discussion follows in section 5.

## 2  Dynamics of large–scale cosmic fields

We assume a universe with vanishing cosmological constant and arbitrary density parameter $\Omega$. The content of the universe is assumed to behave as a pressureless fluid, which undergoes gravitational evolution described by the Newtonian equations

$$\nabla^2 \phi = 4\pi G \rho_b a^2 \delta,$$

$$\frac{\partial}{\partial t}\mathbf{v} + \frac{1}{a}(\mathbf{v}\cdot\nabla)\mathbf{v} + \frac{\dot{a}}{a}\mathbf{v} + \frac{1}{a}\nabla\phi = 0, \qquad (1)$$

$$\frac{\partial}{\partial t}\delta + \frac{1}{a}\nabla\cdot(1+\delta)\mathbf{v} = 0,$$

where $\delta = \delta(\mathbf{x},t) = \delta\rho/\rho_b$ and $\mathbf{v} = \mathbf{v}(\mathbf{x},t)$ are the density contrast and the peculiar velocity, respectively, and $\mathbf{x}$ is the Eulerian comoving coordinate. In the following we will characterize the velocity field by the dimensionless scalar

$$\theta(\mathbf{x}) \equiv \frac{1}{H}\nabla\cdot\mathbf{v}(\mathbf{x}), \qquad (2)$$

where $H$ is the Hubble parameter.

The perturbative expansion of the density contrast around the background $\delta = 0$ is

$$\delta = \delta_1 + \delta_2 + \delta_3 + \cdots, \qquad (3)$$

where $\delta_n = \mathcal{O}(\delta_1^n)$. A similar expansion may be performed for the velocity divergence field $\theta$. The $n$–th order solutions are obtained from equations (1) using the solutions of the $(n-1)$–th order of density and velocity fields as the source terms.

In an Einstein–de Sitter universe, the scale factor $a \propto t^{2/3}$, and the background density $\rho_b = 3\dot{a}^2/8\pi G a^2 = 1/6\pi G t^2$. The time dependence of the $n$–th order follows

$$\delta_n(\mathbf{x},t) = [D(t)]^n \delta_n(\mathbf{x}), \qquad (4)$$

where $D(t) \propto a(t)$ and we consider only the mode growing in time. For an arbitrary cosmological model, however, the time dependences of different orders should be considered independently, i.e. $\delta_n(\mathbf{x},t) = D_n(t)\delta_n(\mathbf{x})$. Fortunately, for a wide range of $\Omega$ the solutions for the density contrast are very weakly dependent on $\Omega$ (see the Appendix) and the E–dS case provides an excellent approximation. For the velocity divergence field, the only strong dependence on $\Omega$ enters via the factor

$$f(\Omega) \equiv \frac{a}{D}\frac{\mathrm{d}D}{\mathrm{d}a} \approx \Omega^{0.6}, \qquad (5)$$

(Peebles 1980), where $D = D_1$ is the time dependence of the first order solution.

All of the following calculations are much simpler if they are performed in Fourier space. For the first order of the density contrast field we have

$$\delta_1(\mathbf{k},t) = D(t)\int \mathrm{d}^3x\ \delta_1(\mathbf{x})e^{-i\mathbf{k}\cdot\mathbf{x}}, \qquad (6)$$

and the inverse Fourier transform is

$$\delta_1(\mathbf{x},t) = D(t)(2\pi)^{-3}\int \mathrm{d}^3k\ \delta_1(\mathbf{k})e^{i\mathbf{k}\cdot\mathbf{x}}. \qquad (7)$$

The first order of the velocity divergence is of the form
$$\theta_1(\mathbf{x}, t) = -f(\Omega)D(t)\delta_1(\mathbf{x}). \tag{8}$$

For the calculation of kurtosis (the fourth reduced moment), only the second– and third–order solutions are needed, and we give them here in the Fourier representation (e.g. Juszkiewicz et al. 1984; Goroff et al. 1986). For the density field we have

$$\delta_2(\mathbf{k}, t) = D^2(t)\frac{1}{(2\pi)^3}\int d^3p \int d^3q \; \delta^3(\mathbf{p}+\mathbf{q}-\mathbf{k})\delta_1(\mathbf{p})\delta_1(\mathbf{q})P_2^{(s)}(\mathbf{p}, \mathbf{q}), \tag{9}$$

$$\delta_3(\mathbf{k}, t) = D^3(t)\frac{1}{(2\pi)^6}\int d^3p \int d^3q \int d^3r \; \delta^3(\mathbf{p}+\mathbf{q}+\mathbf{r}-\mathbf{k})\delta_1(\mathbf{p})\delta_1(\mathbf{q})\delta_1(\mathbf{r})P_3^{(s)}(\mathbf{p}, \mathbf{q}, \mathbf{r}), \tag{10}$$

and for the velocity divergence the solutions are

$$\theta_2(\mathbf{k}, t) = -f(\Omega)D^2(t)\frac{1}{(2\pi)^3}\int d^3p \int d^3q \; \delta^3(\mathbf{p}+\mathbf{q}-\mathbf{k})\delta_1(\mathbf{p})\delta_1(\mathbf{q})P_{2\theta}^{(s)}(\mathbf{p}, \mathbf{q}), \tag{11}$$

$$\theta_3(\mathbf{k}, t) = -f(\Omega)D^3(t)\frac{1}{(2\pi)^6}\int d^3p \int d^3q \int d^3r \; \delta^3(\mathbf{p}+\mathbf{q}+\mathbf{r}-\mathbf{k})\delta_1(\mathbf{p})\delta_1(\mathbf{q})\delta_1(\mathbf{r})P_{3\theta}^{(s)}(\mathbf{p}, \mathbf{q}, \mathbf{r}). \tag{12}$$

The symmetrized kernels are of the form

$$P_2^{(s)}(\mathbf{p}, \mathbf{q}) = \frac{1}{14}J(\mathbf{p}+\mathbf{q}, \mathbf{p}, \mathbf{q}), \tag{13}$$

$$P_{2\theta}^{(s)}(\mathbf{p}, \mathbf{q}) = \frac{1}{14}L(\mathbf{p}+\mathbf{q}, \mathbf{p}, \mathbf{q}), \tag{14}$$

$$\begin{aligned}
P_3^{(s)}(\mathbf{p}, \mathbf{q}, \mathbf{r}) = \; & A\,[\; H(\mathbf{p}+\mathbf{q}+\mathbf{r}, \mathbf{p})J(\mathbf{q}+\mathbf{r}, \mathbf{q}, \mathbf{r}) + \\
& + \; H(\mathbf{p}+\mathbf{q}+\mathbf{r}, \mathbf{q})J(\mathbf{p}+\mathbf{r}, \mathbf{p}, \mathbf{r}) + \\
& + \; H(\mathbf{p}+\mathbf{q}+\mathbf{r}, \mathbf{r})J(\mathbf{p}+\mathbf{q}, \mathbf{p}, \mathbf{q}) + \\
& + \; H(\mathbf{p}+\mathbf{q}+\mathbf{r}, \mathbf{q}+\mathbf{r})L(\mathbf{q}+\mathbf{r}, \mathbf{q}, \mathbf{r}) + \\
& + \; H(\mathbf{p}+\mathbf{q}+\mathbf{r}, \mathbf{p}+\mathbf{r})L(\mathbf{p}+\mathbf{r}, \mathbf{p}, \mathbf{r}) + \\
& + \; H(\mathbf{p}+\mathbf{q}+\mathbf{r}, \mathbf{p}+\mathbf{q})\,L(\mathbf{p}+\mathbf{q}, \mathbf{p}, \mathbf{q})\,] + \\
+ B\,[\; & F(\mathbf{p}+\mathbf{q}+\mathbf{r}, \mathbf{p}, \mathbf{q}+\mathbf{r})L(\mathbf{q}+\mathbf{r}, \mathbf{q}, \mathbf{r}) + \\
& + \; F(\mathbf{p}+\mathbf{q}+\mathbf{r}, \mathbf{q}, \mathbf{p}+\mathbf{r})L(\mathbf{p}+\mathbf{r}, \mathbf{p}, \mathbf{r}) + \\
& + \; F(\mathbf{p}+\mathbf{q}+\mathbf{r}, \mathbf{r}, \mathbf{p}+\mathbf{q})\,L(\mathbf{p}+\mathbf{q}, \mathbf{p}, \mathbf{q})\,],
\end{aligned} \tag{15}$$

where $A = 1/108$ and $B = 1/189$. To obtain the similar expression for $P_{3\theta}^{(s)}$, the constants $A, B$ should be replaced by $A_\theta = 1/252$ and $B_\theta = 1/63$. In the expression above, the notation follows that of Makino et al. (1992), i.e.

$$H(\mathbf{p}, \mathbf{q}) = \frac{\mathbf{p}\cdot\mathbf{q}}{q^2}, \tag{16}$$

$$F(\mathbf{p}+\mathbf{q}, \mathbf{p}, \mathbf{q}) = \frac{1}{2}\frac{|\mathbf{p}+\mathbf{q}|^2\,\mathbf{p}\cdot\mathbf{q}}{p^2 q^2}, \tag{17}$$

$$\begin{aligned}
J(\mathbf{p}+\mathbf{q}, \mathbf{p}, \mathbf{q}) &= 4F(\mathbf{p}+\mathbf{q}, \mathbf{p}, \mathbf{q}) + 5H(\mathbf{p}+\mathbf{q}, \mathbf{p}) + 5H(\mathbf{p}+\mathbf{q}, \mathbf{q}) = \\
&= 4\frac{(\mathbf{p}\cdot\mathbf{q})^2}{p^2 q^2} + 7\frac{p^2+q^2}{p^2 q^2}\mathbf{p}\cdot\mathbf{q} + 10,
\end{aligned} \tag{18}$$

$$\begin{aligned}
L(\mathbf{p}+\mathbf{q}, \mathbf{p}, \mathbf{q}) &= 8F(\mathbf{p}+\mathbf{q}, \mathbf{p}, \mathbf{q}) + 3H(\mathbf{p}+\mathbf{q}, \mathbf{p}) + 3H(\mathbf{p}+\mathbf{q}, \mathbf{q}) = \\
&= 8\frac{(\mathbf{p}\cdot\mathbf{q})^2}{p^2 q^2} + 7\frac{p^2+q^2}{p^2 q^2}\mathbf{p}\cdot\mathbf{q} + 6.
\end{aligned} \tag{19}$$

The smoothing of the fields is introduced by the convolution of the density contrast (or the velocity divergence) with the filtering function,

$$\delta_R(\mathbf{x}, t) = \int \mathrm{d}^3 y \, \delta(\mathbf{y}, t) W_R(|\mathbf{x} - \mathbf{y}|), \tag{20}$$

where the window function is normalized so that $\int \mathrm{d}^3 x W(x) = 1$. We perform our calculations for a Gaussian window function, with Fourier representation

$$W(pR) = \mathrm{e}^{-p^2 R^2 / 2}. \tag{21}$$

## 3 Statistics of density and velocity fields

We assume a Gaussian distribution for $\delta_1$ in (3), and we define

$$\sigma^2 = \langle \delta_1^2 \rangle = D^2(t) \int \frac{\mathrm{d}^3 k}{(2\pi)^3} P(k) W^2(kR) \tag{22}$$

to be the variance of the density field. Its equivalent for the velocity field, $\sigma_\theta^2$, is obtained from the above expression by multiplying by the factor $f^2(\Omega)$. The quantities $\sigma$ and $\sigma_\theta$ are the relevant "smallness" parameters that should control the accuracy of the perturbative approximation. In the case of Gaussian models, the statistical properties of $\delta_1$, and hence of all the following terms in the perturbative series (3), are determined entirely by the power spectrum $P(k)$, which is defined by the relation

$$\langle \delta(\mathbf{k}) \delta(\mathbf{p}) \rangle = (2\pi)^3 \delta^3(\mathbf{k} + \mathbf{p}) P(\mathbf{k}), \tag{23}$$

where $\mathbf{k}, \mathbf{p}$ are comoving wavevectors. We consider initial power spectra with a power–law form,

$$P(k) = k^n, \quad -3 \leq n \leq 1. \tag{24}$$

The deviation of the dynamically evolving fields from a Gaussian distribution can be described by the normalized cumulants,

$$S_l = \frac{M_l}{\sigma^{2l-2}}, \tag{25}$$

where the cumulants (reduced moments) themselves are given by

$$M_l = \frac{\mathrm{d}^l \ln \langle \mathrm{e}^{t\delta} \rangle}{\mathrm{d} t^l} \bigg|_{t=0}. \tag{26}$$

The cumulants can be expressed in terms of the central moments of the distribution; in particular, $M_3 = \langle \delta^3 \rangle$ and $M_4 = \langle \delta^4 \rangle - 3 \langle \delta^2 \rangle^2$.

The significance of the normalized cumulants defined by equation (25) is that they are independent of $\sigma$ to the lowest non-vanishing order in perturbation theory (see Fry 1984; Bernardeau 1992; Juszkiewicz et al. 1993b, hereafter JWACB). In the following, we will refer to $S_3$ and $S_4$ as the skewness and the kurtosis respectively. Conventional statistical usage defines the skewness and kurtosis as $M_3/\sigma^3$ and $M_4/\sigma^4$, respectively, but for gravitational dynamics with Gaussian initial conditions the combinations $M_3/\sigma^4$ and $M_4/\sigma^6$ are more physically relevant, so we can save many words with this slight abuse of terminology.

## 3.1 The skewness

Following the perturbative expansion (3), we have for the density field

$$\langle \delta^3 \rangle = \langle \delta_1^3 \rangle + 3 \langle \delta_1^2 \delta_2 \rangle + \mathcal{O}(\sigma^6). \tag{27}$$

For the Gaussian distribution $\langle \delta_1^3 \rangle = 0$ and the third normalized cumulant (the skewness) is therefore

$$S_3 = \frac{3 \langle \delta_1^2 \delta_2 \rangle}{\sigma^4}. \tag{28}$$

Equations (22) and (24) imply

$$\sigma^2 = D^2(t) \frac{\Gamma(\frac{n+3}{2})}{(2\pi)^2 R^{n+3}}, \tag{29}$$

and by combining with equations (28) and (9) we obtain

$$S_3 = \frac{3}{28\pi^2 \Gamma^2(\frac{n+3}{2})} \int d^3p \int d^3q \, W(p)W(q)W(|\mathbf{p}+\mathbf{q}|)P(p)P(q)J(\mathbf{p}+\mathbf{q},\mathbf{p},\mathbf{q}). \tag{30}$$

The integration over angular variables may be performed using the expansion in Legendre functions of the exponential terms that come in from the window functions

$$e^{-\mathbf{p} \cdot \mathbf{q}} = e^{-pq\mu} = \sum_{m=0}^{\infty} (-1)^m (2m+1) \sqrt{\frac{\pi}{2pq}} I_{m+\frac{1}{2}}(pq) P_m(\mu), \tag{31}$$

where $I_{m+1/2}(pq)$ are the Bessel functions. The function $J$ defined in (18) is easily expressed in terms of the Legendre polynomials $P_m(\mu)$, and the integration becomes as simple as the orthogonality relation

$$\int_{-1}^{1} d\mu \, P_m(\mu) P_n(\mu) = \frac{2\delta_{mn}}{2m+1}. \tag{32}$$

The result is

$$S_3 = \frac{12\sqrt{\pi}}{\sqrt{2} \Gamma^2(\frac{n+3}{2})} \int dp \int dq \, (pq)^{n+3/2} e^{-p^2-q^2} \times$$
$$\times \left[ \frac{34}{21} I_{\frac{1}{2}}(pq) - \left( \frac{p}{q} + \frac{q}{p} \right) I_{\frac{3}{2}}(pq) + \frac{8}{21} I_{\frac{5}{2}}(pq) \right], \tag{33}$$

which can be evaluated using the following expansion of the Bessel functions,

$$I_\nu(z) = \sum_{m=0}^{\infty} \frac{1}{m! \Gamma(\nu+m+1)} \left( \frac{z}{2} \right)^{\nu+2m}, \tag{34}$$

and integrating term by term. Since

$$\int_0^\infty q^\alpha e^{-q^2} dq = \frac{1}{2} \Gamma\left( \frac{\alpha+1}{2} \right), \tag{35}$$

this expansion leads to a series of products of gamma functions, which can in turn be rewritten in terms of the hypergeometric functions, defined by

$$_2F_1(a,b,c,x) = \sum_{m=0}^{\infty} \frac{(a)_m (b)_m}{(c)_m} x^m, \tag{36}$$

and
$$(a)_k \equiv \frac{\Gamma(a+k)}{\Gamma(a)}. \tag{37}$$

The final result, giving the dependence of the skewness on the spectral index, $n$, is

$$S_3 = 3 \; _2F_1\left(\frac{n+3}{2}, \frac{n+3}{2}, \frac{3}{2}, \frac{1}{4}\right) - \left(n + \frac{8}{7}\right) \; _2F_1\left(\frac{n+3}{2}, \frac{n+3}{2}, \frac{5}{2}, \frac{1}{4}\right). \tag{38}$$

Proceeding in exactly the same way, we may calculate the corresponding skewness parameter for the velocity divergence field, and we obtain

$$S_{3\theta} = -\frac{1}{f(\Omega)}\left[3 \; _2F_1\left(\frac{n+3}{2}, \frac{n+3}{2}, \frac{3}{2}, \frac{1}{4}\right) - \left(n + \frac{16}{7}\right) \; _2F_1\left(\frac{n+3}{2}, \frac{n+3}{2}, \frac{5}{2}, \frac{1}{4}\right)\right]. \tag{39}$$

In this case the dependence on $\Omega$ is not negligible. With the above formulas we easily recover the values of the skewness obtained by JBC, while avoiding their tedious calculations and generalizing the results to arbitrary (non-integer) $n$. An expression equivalent to equation (38) was recently obtained independently by Matsubara (1994).

## 3.2 The kurtosis

For the fourth moment of the density contrast distribution, the perturbative expansion (3) gives

$$\langle \delta^4 \rangle = \langle \delta_1^4 \rangle + 6\langle \delta_1^2 \delta_2^2 \rangle + 4\langle \delta_1^3 \delta_3 \rangle + \mathcal{O}(\sigma^8), \tag{40}$$

where $\langle \delta_1^4 \rangle = 3\sigma^4$, and the fourth normalized cumulant (the kurtosis) is

$$S_4 = \frac{\langle \delta^4 \rangle - 3\sigma^4}{\sigma^6} = \frac{6\langle \delta_1^2 \delta_2^2 \rangle}{\sigma^6} + \frac{4\langle \delta_1^3 \delta_3 \rangle}{\sigma^6} = I_1 + I_2. \tag{41}$$

Only the connected parts of the tree diagrams are used for the calculation (see Fry 1984, Bernardeau 1992). The required integrals are

$$\begin{aligned}
I_1 &= \frac{3}{98\pi^3 \Gamma^3(\frac{n+3}{2})} \int d^3p \int d^3q \int d^3r \; P(p)P(q)P(r) \times \\
&\times W(|\mathbf{p}+\mathbf{q}|)W(|\mathbf{r}-\mathbf{q}|)W(p)W(r) \times \\
&\times J(\mathbf{p}+\mathbf{q},\mathbf{p},\mathbf{q})J(\mathbf{r}-\mathbf{q},\mathbf{r},-\mathbf{q}),
\end{aligned} \tag{42}$$

$$\begin{aligned}
I_2 &= \frac{9}{\pi^3 \Gamma^3(\frac{n+3}{2})} \int d^3p \int d^3q \int d^3r \; P(p)P(q)P(r) \times \\
&\times W(|\mathbf{p}+\mathbf{q}+\mathbf{r}|)W(p)W(q)W(r) \times \\
&\times [\, A\,[\; H(\mathbf{p}+\mathbf{q}+\mathbf{r},\mathbf{r})J(\mathbf{p}+\mathbf{q},\mathbf{p},\mathbf{q}) + \\
&\quad + H(\mathbf{p}+\mathbf{q}+\mathbf{r},\mathbf{p}+\mathbf{q})L(\mathbf{p}+\mathbf{q},\mathbf{p},\mathbf{q})] + \\
&\quad + B\,[\; F(\mathbf{p}+\mathbf{q}+\mathbf{r},\mathbf{r},\mathbf{p}+\mathbf{q}) \; L(\mathbf{p}+\mathbf{q},\mathbf{p},\mathbf{q})]\,].
\end{aligned} \tag{43}$$

Again we can use the expansion (31) and integrate with respect to the angular variables using the orthogonality relation (32) to get

$$\begin{aligned}
I_1 &= \frac{48\pi}{\Gamma^3(\frac{n+3}{2})} \int dp \int dq \int dr \; (pr)^{n+3/2} q^{n+1} e^{-p^2-q^2-r^2} \times \\
&\times \left[\frac{34}{21} I_{\frac{1}{2}}(pq) - \left(\frac{p}{q}+\frac{q}{p}\right) I_{\frac{3}{2}}(pq) + \frac{8}{21} I_{\frac{5}{2}}(pq)\right] \times \\
&\times \left[\frac{34}{21} I_{\frac{1}{2}}(qr) - \left(\frac{q}{r}+\frac{r}{q}\right) I_{\frac{3}{2}}(qr) + \frac{8}{21} I_{\frac{5}{2}}(qr)\right].
\end{aligned} \tag{44}$$

As before, the expression is integrable analytically, and the resulting series of gamma functions converges fast enough to be summed numerically to any desired accuracy.

In the case of $I_2$, we need to change variables into $\mathbf{l} = \mathbf{p} + \mathbf{q}$ (and in the following we use $\mathbf{l} \cdot \mathbf{p} = lp\beta$), and the method for the angular integration works only partly, to produce

$$\begin{aligned}
I_2 &= \frac{144\sqrt{2\pi}}{\Gamma^3(\frac{n+3}{2})} \int dp \int dl \int dr \, l^2 (pr)^{n+2} e^{-l^2 - p^2 - r^2} \times \\
&\times \left\{ \left[ (A + \frac{B}{3}) r^{-\frac{1}{2}} l^{-\frac{1}{2}} I_{\frac{1}{2}}(rl) - (A + \frac{B}{2}) r^{\frac{1}{2}} l^{-\frac{3}{2}} I_{\frac{3}{2}}(rl) + \right. \right. \\
&\quad \left. - \frac{B}{3} r^{-\frac{3}{2}} l^{\frac{1}{2}} I_{\frac{3}{2}}(rl) + \frac{2B}{3} r^{-\frac{1}{2}} l^{-\frac{1}{2}} I_{\frac{5}{2}}(rl) \right] \times \\
&\times \int_{-1}^{1} d\beta \, L(l, p, \beta)(l^2 + p^2 - 2lp\beta)^{\frac{n}{2}} e^{lp\beta} + \\
&\quad + A \left[ r^{-\frac{1}{2}} l^{-\frac{1}{2}} I_{\frac{1}{2}}(rl) - r^{\frac{1}{2}} l^{-\frac{3}{2}} I_{\frac{3}{2}}(rl) \right] \times \\
&\times \left. \int_{-1}^{1} d\beta \, J(l, p, \beta)(l^2 + p^2 - 2lp\beta)^{\frac{n}{2}} e^{lp\beta} \right\}.
\end{aligned} \quad (45)$$

A complicated but analytic expression for the integrals with respect to $r$ and $\beta$ can be obtained using the MATHEMATICA package. We are left with 2–dimensional integrals over $p$ and $l$, to be performed numerically.

The kurtosis of the velocity divergence field, $S_{4\theta}$, is obtained from the above expressions by replacing the functions $J$ in $I_1$ by $L$ defined in (19), and the constants $A, B$ in $I_2$ by $A_\theta$ and $B_\theta$ given after equation (15). The whole expression must then be multiplied by the factor $1/f^2(\Omega)$.

The integrals (44) and (45) diverge logarithmically for $n = -3$. However, divergences of the same kind come from $\sigma^6$, and in the limit of small $p, q, r$ we may put the window functions $W(k) = 1$, which corresponds to the unsmoothed case, and obtain the remarkable values of $S_4 = 60712/1323 = 45.9$ and $S_{4\theta} = 12088/441 = 27.4$.

## 4  N–body results

There are several reasons for comparing analytic calculations of skewness and kurtosis to results obtained from cosmological N–body simulations. First, this comparison allows us to test the fundamental assumption of our calculations, that it is valid to compute smoothed large-scale quantities from perturbation theory even when the density field is strongly non-linear on small scales. While this assumption seems eminently plausible, we do not know of any rigorous analytic way to demonstrate its validity. Second, the N–body simulations can indicate the range of $\sigma$ over which the perturbative results are accurate. Third, the comparison provides a check against possible errors in the (rather complicated) analytic computations. Finally, the comparison allows one to check the accuracy of the N–body method in the weakly non–linear regime.

It may seem circular to suggest that N–body simulations can check the analytic results and that the analytic results can simultaneously check the N–body methods. However, the analytic and numerical approaches are completely different, so when they yield identical results it seems unlikely that they could both be making mistakes and arriving at the same wrong answer. It thus seems reasonable to interpret agreement as a confirmation of both calculations. Interpretation is much trickier when the N–body and analytic results do not agree perfectly; we will encounter an instance of this problem shortly.

JWACB computed the skewness and kurtosis of density and velocity fields taken from N–body simulations with an $n = -1$ initial power spectrum. For the purpose of this paper, we have run additional simulations with $n = 0$ and $n = +1$ initial spectra. The numerical parameters of the new simulations are similar to those of the $n = -1$ simulations reported in JWACB. We refer the reader to that paper for a detailed description of the simulations; here we provide a brief summary.

All of the simulations use a particle–mesh N–body code written by C. Park (Park 1990, 1991). There are eight independent simulations for each initial spectrum. Each simulation uses a $200^3$ staggered mesh for force computations, evolving particles from expansion factor $a = 1/128$ to the final expansion factor $a = 1$. Evolved density fields are computed on a $100^3$ grid at expansion factors $a = 1/8$, $1/4$, $1/2$, and 1. Moments of these density fields are computed for Gaussian smoothing lengths of $L/50$, $L/25$, and $L/12.5$, where $L = 100$ cells is the size of the box. The $n = -1$ and $n = 0$ simulations use $100^3$ particles. We also ran a full set of $n = +1$ simulations with $100^3$ particles, but after finding small discrepancies with the perturbative results for this spectrum, we repeated the $n = +1$ simulations with $200^3$ particles, still using a $200^3$ force mesh. These higher density simulations yield slightly (but only slightly) better agreement with perturbation theory, and for $n = +1$ it is these results that we report in this paper.

Figures 1 and 2 show, respectively, the values of $S_3$ and $S_{3\theta}$ obtained from the N–body density and velocity divergence fields, plotted for the three initial power spectra as a function of the r.m.s. fluctuations $\sigma$ and $\sigma_\theta$. Squares, triangles, and circles represent results for smoothing lengths $L/12.5$, $L/25$, and $L/50$, respectively. Figures 3 and 4 present the values of $S_4$ and $S_{4\theta}$, with triangles and circles corresponding to $L/25$ and $L/50$ smoothing lengths respectively. With smoothing length $L/12.5$, there are too few independent smoothing volumes to yield statistically useful measures of the kurtosis. In every panel, points show the mean value of $S_p$ from the eight simulations. Error bars indicate 1–$\sigma$ *statistical* uncertainties in this mean value, obtained by taking the dispersion of $S_p$ values from the $N_{sim} = 8$ simulations and dividing by $\sqrt{N_{sim} - 1} = \sqrt{7}$. Solid lines show the numerical results of the perturbative calculations, which are expected to apply in the limit of low $\sigma$. For the $n = -1$ power spectrum, the results in these Figures are identical to those shown in JWACB. Also, while we have included the $S_3$ and $S_{3\theta}$ results for completeness, the new analytic results reported in this paper are for kurtosis, and it is on this comparison that we focus most of the following discussion.

Statistical error bars on the N–body points are smaller for the shorter smoothing lengths because there are more independent smoothing volumes per simulation. However, it is important to note that there may be systematic errors in the N–body skewness and kurtosis values because of the numerical limitations of the simulations themselves; recall that these quantities are obtained from the ratios of two dimensionless numbers, each of order $\sigma^4$ in the case of skewness and $\sigma^6$ in the case of kurtosis. The systematic errors are probably larger for shorter smoothing lengths, since results at this scale are more sensitive to the finite force resolution of the simulations. Another potential error comes from the cut–off of long waves at the box size, and this error is larger for larger smoothing lengths. However, we suspect that this possible systematic effect is not important at the smoothing lengths we use, in part because $L/25$ is much smaller than the box size $L$, and in part because we obtain the best agreement between the N–body and perturbative $S_4$ results for $n = -1$, the spectrum with the most large–scale power.

In the case of density fields, for the spectral index $n = -1$, the N–body and perturbative results agree perfectly at low values of $\sigma$. The rise in $S_4$ at $\sigma \approx 1$ probably reflects the breakdown of perturbation theory in this regime, though even at $\sigma = 2$ the kurtosis is within a factor of 2 of the value predicted from perturbation theory. For $n = 0$ the N–body results lie slightly below the perturbative values at low $\sigma$, though within the error bars of the smoothing length $L/25$ points. For $n = +1$ the discrepancy is somewhat larger; agreement within the error bars occurs only for $\sigma \lesssim 0.1$, where the N–body errors are rather large. A possible explanation of the larger apparent discrepancy is that for this spectrum the perturbative approximation becomes (mildly) inaccurate at a relatively low value of $\sigma$. However, the discrepancy for $0.1 \lesssim \sigma \lesssim 0.5$ might also arise partly from inaccuracies in the N–body simulations themselves. In particular, because the $n = +1$ spectrum has more small-scale power than the other spectra, simulations with higher force resolution may be needed in order to obtain a similar level of accuracy. The skewness results (lower panel of Figure 1) are similar to the kurtosis results; N–body and perturbative calculations agree for $\sigma \lesssim 0.1$, where the N–body error bars are rather large, but the N–body simulations yield systematically lower values ($S_3 \approx 2.8$ vs. the analytic value of 3.03) when $\sigma > 0.1$.

While short of perfect, the agreement between the N–body and perturbative calculations of density–field moments is remarkably good. For the velocity divergence field, the situation is less satisfying. Skewness values agree reasonably well (Figure 2), at least at the larger smoothing lengths, though there are significant discrepancies in the $n = -1$ case for smoothing length $L/25$ and $\sigma_\theta \gtrsim 0.25$. For the kurtosis, however, agreement is quite poor, especially for $n = -1$ (Figure 4). We can think of at least four possible causes of this discrepancy: (1) there is an error in the analytic calculations, (2) the perturbative approximation for this quantity has already become inaccurate by the time $\sigma_\theta \approx 0.1$, (3) the N–body simulations do not evolve the velocity divergence field with sufficient accuracy, (4) the definition of the smoothed velocity divergence from the final conditions of the N–body simulation is too noisy to allow accurate measurement of the kurtosis.

We regard explanation (1) as highly unlikely, since the analytic computation of $S_{4\theta}$ is a straightforward generalization of the computation of $S_4$, and in the density case we find good agreement with the simulations. Explanation (2) is somewhat more plausible; perturbative and N–body results for the density kurtosis agree well for $\sigma$ as large as $1/2$, but the perturbative approximation need not have the same range of validity for the density field and for the velocity divergence. For the skewness (Figure 2), there is a weak trend towards better agreement with perturbation theory at the lowest values of $\sigma_\theta$, at least for the $n = -1$ spectrum where the discrepancy at $S_{4\theta}$ is most vexing. A restricted range of validity could make it quite difficult to demonstrate concordance between perturbation theory and N–body simulations, since the former would be accurate only in the regime where the statistical errors of the simulations are large. Nonetheless, it seems unlikely that a breakdown of perturbation theory can explain all of the discrepancies in Figure 4, since $\sigma_\theta \approx 0.1$ is a quite low value of the relevant "smallness" parameter, and since the agreement between perturbation theory and the mean N–body value of $S_{4\theta}$ does not improve as one goes to lower values of $\sigma_\theta$.

Errors in the N–body results could arise either because the gravitational evolution of the N–body particle distribution is incorrect (explanation 3) or because the measurement of the velocity divergence from the discrete particle distribution is noisy (explanation 4). Since the N–body density fields agree well with perturbation theory, we have no reason to expect that the particle velocities are incorrect. However, displacements involve double integrals of gravitational accelerations, while velocities involve only single integrals, so it is possible that discreteness effects and the force errors that arise from computations on a finite mesh influence the velocity divergence more than they influence the density. Errors in the definition of the smoothed velocity field could be important even if the particle velocities themselves are essentially perfect, especially in low-density regions, where there may be too few particles to yield accurate velocity estimates.

As detailed in JWACB, we create our smoothed velocity fields in two stages, a mass–weighted smoothing onto a cubic grid using a small–scale Gaussian filter, followed by a volume–weighted smoothing using a larger filter. In tests with top–hat smoothing, Bernardeau (private communication) finds that the probability distribution of the velocity divergence is quite sensitive to the method used for defining the field. Gaussian smoothing should be somewhat more stable than top–hat smoothing in this regard, but perhaps not stable enough to suppress all noise in the measured velocity divergence.

We suspect that the order in which we have listed our possible explanations is in fact the order of increasing importance, i.e. that item (1) does not arise, that item (2) affects the comparison at moderate $\sigma_\theta$ but not at the lowest values, and that items (3) and (4) are the main contributors to the discrepancies seen in Figure 4, with item (4) the most important. However, without N–body results that *do* match the perturbative calculations convincingly at low $\sigma_\theta$, we cannot draw any firm conclusions on this point.

## 5 Discussion

### 5.1 Results

Tables 1 and 2 and Figures 5–8 summarize the results of the preceding Sections. Table 1 lists skewness values for the density contrast and velocity divergence from perturbation theory (equations (38) and (39)) and from our N–body calculations. The perturbative values of $S_3$ and $S_{3\theta}$ at integer values of $n$ are identical to those first reported by JBC and Bernardeau et al. (1994), respectively. Table 2 lists perturbative and N–body values for the kurtosis. Figures 5–8 plot these results, with filled circles showing the perturbative values, and open triangles showing the N–body estimates.

Our condensation of the N–body results shown in Figures 1–4 into a single value and error bar for each spectral index requires some rather arbitrary choices. We have decided to compute the values in Tables 1 and 2 and Figures 5–8 by taking an unweighted average of all of the smoothing length $L/25$ points for which the r.m.s. fluctuation lies in the range $-1.2 < \log_{10}\sigma < -0.5$ (or $\sigma_\theta$ for the velocity divergence). The $L/50$ results have smaller statistical error bars, but we have not used them because they are more susceptible to systematic errors arising from the finite force resolution of the N–body simulations. We have not used points with higher values of $\sigma$ because the perturbative approximation is more likely to break down in this regime, and we are interested in seeing how well perturbation theory and N–body simulations agree in the range where they ought to yield consistent results. We have used an unweighted rather than a weighted fit because we do not want to assign most of the weight to the point with the highest value of $\sigma$ (and lowest statistical error), as this might again tilt the comparison to a regime where perturbation theory is breaking down. Finally, we have eliminated the lowest values of $\sigma$, because for these the statistical error bars are very large. Our strategy seems a reasonable compromise between competing concerns, but readers should not take the summary N–body values and error bars in these Tables too seriously; the more detailed comparison between N–body results and perturbation theory in Figures 1–4 is far more meaningful.

The fourth column of Table 2 and the open squares in Figure 3 show estimates of $S_4$ obtained from Monte Carlo integration by Catelan & Moscardini (1994). Our perturbative results lie outside their quoted uncertainties for spectral indices $n > -1$. For $n = -1$ and $n = 0$, the N–body results are consistent with either our perturbative calculations or the Monte Carlo integrations, to within our rather conservative estimates of the N–body error bars. For $n = +1$, the N–body simulations are marginally consistent with our perturbative calculation, and they are inconsistent with the Monte Carlo results. In addition to this better agreement with N–body simulations, we have two further reasons for believing that our results are more accurate than the Monte Carlo results. First, although our calculation requires a numerical integration, this integral is much more straightforward from a numerical point of view than the one performed by Catelan & Moscardini (1994). Monte Carlo integrations are difficult when the integrand is strongly peaked in rather small regions of a multi-dimensional space. Furthermore, our numerical integration is more reliable for the *higher* spectral indices, as the integrand is less variable and the integration converges more quickly. Second, we can compare our results for $n = +1$, the case of largest discrepancy, to the results of an independent Monte Carlo integration, that of Goroff et al. (1986). They analyze a cold dark matter spectrum, but at large smoothing radii the index of this spectrum *is* $n = +1$. They obtain $S_4 = 16 \pm 1$, in perfect agreement with our value of 15.95. On the scale where the effective index of the CDM spectrum is $n \approx -1$, Goroff et al. find $S_4 \approx 20$, in good agreement with our results from perturbation theory, our N–body estimates, and Catelan & Moscardini's Monte Carlo computations.

The fifth column of Table 2 and the filled circles of Figure 8 show perturbative results for $S_{4\theta}$ when $\Omega = 1$. The corresponding N–body results are shown in the last column of Table 2 and by open triangles in Figure 8. We find acceptable agreement for $n = +1$ and $n = 0$, but significant disagreement for $n = -1$. We have discussed possible reasons for this discrepancy in section 4. Unfortunately, we are not aware of any independent estimates of $S_{4\theta}$ (e.g. from Monte Carlo integrations).

## 5.2 Comparison with top-hat filtering

As a more roundabout check of our results, we can use a heuristic technique proposed by Bernardeau (1994). For power–law initial spectra and a *top–hat* smoothing filter, analytical expressions are known for arbitrary normalized cumulant $S_p$, e.g.

$$S_3 = \frac{34}{7} - (n+3), \tag{46}$$

(JBC), and

$$S_4 = \frac{60712}{1323} - \frac{62}{3}(n+3) + \frac{7}{3}(n+3)^2, \tag{47}$$

(Bernardeau 1994). We also know $S_3$ in the case of Gaussian smoothing (38), and for any spectral index $n$ we can compute from (46) the corresponding effective index $n_{eff}$ that would yield the same value of $S_3$ under top–hat smoothing. We can then substitute $n_{eff}$ into the expression (47) to estimate the value of $S_4$ for a Gaussian filter. For $n = -1$ we obtain $n_{eff} = -1.61$ and $S_4 = 21.7$ which is in good agreement with our result $S_4 = 21.9$. For other spectral indices in Table 2 the discrepancy is even smaller, especially for the questionable case $n = +1$, where it is only 0.05. This excellent agreement offers another argument for the correctness of our calculations.

For the velocity divergence field smoothed with a *top-hat* filter, the skewness and kurtosis parameters are

$$S_{3\theta} = -\frac{1}{f(\Omega)} \left[ \frac{26}{7} - (n+3) \right], \tag{48}$$

$$S_{4\theta} = \frac{1}{f^2(\Omega)} \left[ \frac{12088}{441} - \frac{338}{21}(n+3) + \frac{7}{3}(n+3)^2 \right], \tag{49}$$

(Bernardeau 1994). We find that the discrepancy between our perturbative results and the values of $S_{4\theta}$ calculated from (49) with $n_{eff}$ grows with the spectral index. For instance (49) with $n_{eff} = -1.48$ reproduces our result for $n = -1$ almost exactly (the difference between them is smaller than 0.01), while for $n = +1$ the discrepancy is of the order of 1. This difference is probably due to the specific features of the $S_{3\theta}$ and $S_{4\theta}$ functions.

In general we notice that the values of skewness and kurtosis decrease with $n$ faster for top-hat smoothing than for Gaussian smoothing. Therefore we always need $n_{eff} \leq n$ to recover the Gaussian smoothing values from the top-hat formulas. The differences between the results for the two filters become larger for larger values of $n$, while both of them yield the same value of $S_p$ in the case of $n = -3$, corresponding to no smoothing. With this motivation in mind, we attempt to find quasi-analytical, simple, polynomial formulas for the $S_p$, similar to those that apply for top-hat filtering:

$$S_p = \sum_k a_k (n+3)^k, \qquad k = 0, 1, 2, \ldots. \tag{50}$$

In the case of skewness the attempt is perfectly justified, as we may expand the hypergeometric functions in expressions (38) and (39) in powers of $n+3$ simply by rearranging the appropriate terms of the hypergeometric series. We find that the first two terms of the expansion (i.e. the free and the linear term) are exactly the same as in the formula for the skewness under top-hat filtering (46) or (48). The higher power terms are fitted numerically, and we see that it is sufficient to add only two more terms to obtain polynomial approximations of expressions (38) and (39) with accuracy of 0.2% or better. We apply the same numerical procedure to the values of kurtosis and notice that the first two terms are again very well fitted by the analogous terms of top-hat formulas (47) and (49). The coefficients of the polynomial approximations for skewness and kurtosis for both the density and velocity divergence fields are given in Table 3. We might feel tempted to express the Gaussian filter kurtosis as the top-hat formulas plus higher power terms. This is however restrained by the fact that the coefficients at the quadratic terms of (47) and (49) do not provide good fits for similar terms in our expansion (50).

## 5.3 Concluding remarks

The investigation of moments of cosmological fields has become a small industry in the last few years, spurred in large part by the appearance of large galaxy redshift surveys, which permit measurements that have interestingly small error bars (e.g. Saunders et al. 1991; Gaztañaga 1992; Bouchet et al. 1993; Gaztañaga 1994). The results of these analyses are generally consistent with the hypothesis that large–scale structure developed by gravitational instability from Gaussian initial conditions. Analyses of cosmic microwave background (CMB) fluctuations detected by the COBE satellite and of large–scale peculiar motions inferred from redshift–independent distance measurements are also consistent with Gaussian initial conditions (Smoot et al. 1994; Kofman et al. 1994). CMB fluctuations and peculiar velocities have the advantage that they respond directly to mass fluctuations, while the relation between galaxy counts and the underlying mass distribution is uncertain. However, the source terms for Sachs–Wolfe CMB fluctuations and for large–scale peculiar velocities both involve an integral over the density contrast $\delta(\mathbf{x})$, and this integration over many independent elements allows the central limit theorem to exert its inexorable grasp. As a result, one can have primordial density fluctuations that are strongly non–Gaussian and nonetheless have CMB fluctuations and peculiar velocity distributions that are almost perfectly Gaussian (Scherrer 1992; Scherrer & Schaefer, in preparation). [Note that the distribution of peculiar *velocities* is different in this regard from the distribution of the peculiar velocity *divergence*, considered in this paper, because the derivatives involved in the divergence undo the smearing effect of the integration over $\delta(\mathbf{x})$.] To the extent that one can solve the problem of "bias" between the galaxy and mass distributions – and to some extent one can, as discussed below – one can obtain much more stringent constraints on primordial non–Gaussianness from galaxy counts.

Approaches based on CMB fluctuations, on peculiar velocity statistics, on reconstruction of initial conditions (Nusser & Dekel 1992, 1993; Weinberg 1992), on galaxy count distributions, and on other clustering measures (Weinberg & Cole 1992) have quite different strengths and defects, so it seems wise to pursue a variety of strategies and see whether they converge on a consistent answer. Most analyses of existing data favor the simple hypothesis of Gaussian initial conditions. Constraints from galaxy counts should become much tighter over the next few years, as measurements from the next generation of large galaxy redshift surveys become available.

The present paper brings to a close one chapter of the cosmic–moments story, namely the analytic computation of low–order moments for Gaussian initial conditions and scale–free initial power spectra. Fry (1984) and Bernardeau (1992) solved this problem in the absence of smoothing; the first computations that correctly incorporated smoothing of the final fields were done for the skewness of the density by JBC and for skewness of the velocity divergence by Bernardeau et al. (1994). Bernardeau (1993) computed the kurtosis for the density and the velocity divergence smoothed with a top–hat filter. Our paper rounds out the program by providing, for the Gaussian–filter case, general analytic expressions for the skewness and semi–analytic computations for the kurtosis of both the density and the velocity divergence.

Bernardeau (1994) has recently exploited remarkable geometrical properties of the top–hat filter to compute the *full* hierarchy of $S_p$ coefficients to lowest non-vanishing order in perturbation theory. While it would be nice to know the full hierarchy of $S_p$'s for a Gaussian filter, the magic trick that Bernardeau (1994) uses for the top–hat filter does not work for a Gaussian, and we are doubtful that an equivalent technique can be found for this case. Given the inevitable limitations of the observational data, there is probably not much reason to continue laborious moment–by–moment calculations for the Gaussian filter to higher order.

There are still many aspects of moments of cosmic fields that deserve analytic and numerical investigation. One is the computation of moments beyond the first non–vanishing order in perturbation theory, to understand the departures from constant $S_p$. To date this program has been pursued only for the variance (in the guise of the power spectrum; Juszkiewicz et al. 1984; Makino et al. 1992; Jain & Bertschinger 1993), and in a rough way for the skewness, via the Zel'dovich approximation (Bernardeau & Kofman 1994).

Another area for investigation is that of non–Gaussian initial conditions; studies of non–Gaussian cases are crucial if we are to understand what models of primordial fluctuations can be ruled out by the concordance of observed moments with the hierarchy predicted for Gaussian initial conditions. There have been a few analytic treatments of the non–Gaussian problem (Scherrer 1992; Luo & Schramm 1993; Fry & Scherrer 1994; Chodorowski, in preparation), but without smoothing of the final fields. The combination of non–Gaussian initial conditions with smoothing makes a very tough problem for analytic techniques, and we do not know of any papers that tackle it. N–body results for some non–Gaussian models appear in Weinberg & Cole (1992), but there have been no systematic numerical studies of the behaviour of non–Gaussian models in the weakly non–linear regime.

A third important problem concerns the effect of "biased" galaxy formation on the moments of the density field. Fry & Gaztañaga (1993), JWACB, and Fry (1994) have studied the case of a non–linear but local relation between the galaxy and mass density fields; they find that the effect of bias can be computed readily in the perturbative regime by taking the appropriate order Taylor expansion of the biasing relation. The important general result is that a local bias preserves the hierarchical form of moment relations predicted for the mass distribution, though the values of coefficients like $S_3$ and $S_4$ change in a way that depends on the biasing scheme. The next problem is to consider bias models that do not obey the local relation perfectly. Frieman & Gaztañaga (1994) have investigated the "cooperative" bias scheme of Bower et al. (1993) and find that it predicts a strong feature in the dependence of $S_3$ on scale, a feature that is not found in existing observational data.

The most significant progress over the next few years should come from improving the comparison between theory and observation. A powerful way to approach this problem is to use the full shape of the 1–point probability distribution function (PDF) instead of moments alone. This approach allows one to use observational data in a statistically efficient way, minimizing the effect of uncertainties in the tails of the distribution, which arise because samples probe limited cosmological volumes. JWACB (and, independently, Bernardeau & Kofman 1994) have shown that the Edgeworth series, an asymptotic expansion of the PDF in powers of $\sigma$, can be used to compute the evolution of the density or velocity divergence PDF in the weakly non–linear regime. Bernardeau (1992, 1994; see also Balian & Schaeffer 1989) describes an alternative method based on the Laplace transform. The Laplace–transform method requires knowledge of the full $S_p$ hierarchy, while successive orders of the Edgeworth expansion require successively higher values of $S_p$. In particular, to compute the Edgeworth expansion up to terms of order $\sigma^2$, one needs to know the moments $S_3$ and $S_4$. This indirect use of $S_4$ in the Edgeworth series or other approximations to the PDF may turn out to be the most important application of the new results in this paper, leading to new observational tests of the hypotheses of gravitational instability and Gaussian primordial fluctuations.

## Acknowledgements

We thank Francis Bernardeau and Michał Chodorowski for helpful discussions and Changbom Park for the use of his N–body code. EŁ and RJ thank Alain Omont for his hospitality at Institut d'Astrophysique de Paris, where part of this work was done. RJ, DW, and FB acknowledge the hospitality of the Aspen Center for Physics, and RJ that of the Institute for Advanced Study, during the final phase of this work. This research has been supported in part by the French Ministry of Research and Technology within the programme RFR, and by the Polish State Committee for Scientific Research grant No. 2-1243-91-01. DW acknowledges support from the W.M. Keck Foundation and from NSF grant PHY92-45317.

# Appendix

In section 2 we mentioned the weak dependence on $\Omega$ of the second– and third–order solutions for density and velocity fields. In the following, we show how these functions affect the skewness parameter. Instead of the functions $J$ and $L$, which define the kernels $P_2^s$ and $P_{2\theta}^s$ respectively (see equations (13)–(19) of section 2), we introduce

$$J = 7\left[(1+K) + (\frac{p}{q}+\frac{q}{p})\mu + (1-K)\mu^2\right], \tag{51}$$

$$L = 7\left[2C + (\frac{p}{q}+\frac{q}{p})\mu + 2(1-C)\mu^2\right], \tag{52}$$

where $\mathbf{p} \cdot \mathbf{q} = pq\mu$ and the $\Omega$–dependent functions are

$$K(\Omega) = \frac{3}{7}\Omega^{-2/63}, \tag{53}$$

$$C(\Omega) = \frac{3}{7}\Omega^{-1/21}. \tag{54}$$

The approximation (53) is accurate to within 0.4% in the range $0.05 < \Omega < 3$ (Bouchet et al. 1992), and (54) is accurate to within 2% for $0.1 < \Omega < 10$ (Bernardeau et al. 1994). The changes, fortunately, do not introduce any trouble in the integration, and the results of section 3 are only slightly distorted,

$$S_3 = 3 \; {}_2F_1\left(\frac{n+3}{2}, \frac{n+3}{2}, \frac{3}{2}, \frac{1}{4}\right) - (n+2-2K) \; {}_2F_1\left(\frac{n+3}{2}, \frac{n+3}{2}, \frac{5}{2}, \frac{1}{4}\right), \tag{55}$$

$$S_{3\theta} = -\frac{1}{f(\Omega)}\left[3 \; {}_2F_1\left(\frac{n+3}{2}, \frac{n+3}{2}, \frac{3}{2}, \frac{1}{4}\right) - (n+4-4C) \; {}_2F_1\left(\frac{n+3}{2}, \frac{n+3}{2}, \frac{5}{2}, \frac{1}{4}\right)\right]. \tag{56}$$

The dependence of the third–order solutions for the density and velocity divergence fields on the value of $\Omega$ was considered by Bernardeau (1993) and more recently by Bouchet et al. (1994). They found functions analogous to (53) and (54), which slightly alter the solutions when compared to the case of $\Omega = 1$. As in the second–order case, the overall shape of the third–order solutions used for the calculation of kurtosis does not change. Since we do not have the exact analytical expression for the kurtosis of the fields smoothed with the Gaussian filter, we have not attempted to take into account the weak dependences on $\Omega$. Their influence would probably be of the order of the accuracy of our results.

| spectral index $n$ | $S_3$ (perturbative) | $S_3$ (N–body) | $|S_{3\theta}|$ (perturbative) | $|S_{3\theta}|$ (N–body) |
|---|---|---|---|---|
| -3.0 | 4.86 | – | 3.71 | – |
| -2.5 | 4.40 | – | 3.25 | – |
| -2.0 | 4.02 | – | 2.85 | – |
| -1.5 | 3.71 | – | 2.50 | – |
| -1.0 | 3.47 | $3.48 \pm 0.23$ | 2.19 | $1.90 \pm 0.22$ |
| -0.5 | 3.28 | – | 1.92 | – |
| 0 | 3.14 | $3.05 \pm 0.28$ | 1.67 | $1.42 \pm 0.25$ |
| 0.5 | 3.06 | – | 1.44 | – |
| 1.0 | 3.03 | $2.84 \pm 0.10$ | 1.22 | $1.16 \pm 0.12$ |

Table 1:

The skewness of density ($S_3$) and velocity divergence ($S_{3\theta}$) fields as functions of the spectral index $n$

| spectral index $n$ | $S_4$ (perturbative) | $S_4$ (N–body) | $S_4$ (Monte–Carlo) | $S_{4\theta}$ (perturbative) | $S_{4\theta}$ (N–body) |
|---|---|---|---|---|---|
| -3.0 | 45.9 | – | – | 27.4 | – |
| -2.5 | 37.0 | – | 35 ± 4 | 20.5 | – |
| -2.0 | 30.4 | – | 28 ± 3 | 15.3 | – |
| -1.5 | 25.5 | – | 23 ± 3 | 11.4 | – |
| -1.0 | 21.9 | 19.4 ± 3.5 | 19.5 ± 2 | 8.31 | 4.8 ± 2.4 |
| -0.5 | 19.3 | – | 16 ± 1 | 5.89 | – |
| 0 | 17.5 | 15.6 ± 5.1 | 14 ± 1 | 3.91 | 3.5 ± 3.7 |
| 0.5 | 16.4 | – | 11 ± 1 | 2.24 | – |
| 1.0 | 15.9 | 14.1 ± 2.0 | 9 ± 1 | 0.795 | 2.3 ± 1.8 |

Table 2:

The kurtosis of density ($S_4$) and velocity divergence ($S_{4\theta}$) fields as functions of the spectral index $n$

| coefficient | $S_3$ | $|S_{3\theta}|$ | $S_4$ | $S_{4\theta}$ |
|---|---|---|---|---|
| $a_0$ | 4.85714 | 3.71428 | 45.8896 | 27.4104 |
| $a_1$ | -1. | -1. | -20.6667 | -16.0952 |
| $a_2$ | 0.168016 | 0.143571 | 6.03051 | 4.71684 |
| $a_3$ | -0.00814026 | -0.0123706 | -1.01031 | -0.850292 |
| $a_4$ | – | – | 0.0817073 | 0.0653419 |
| accuracy | 0.2 % | 0.2 % | 0.15 % | 1.6 % |

Table 3:

Coefficients of the polynomial approximation formulas for the skewness and kurtosis parameters of the density and velocity divergence fields smoothed with a Gaussian filter.

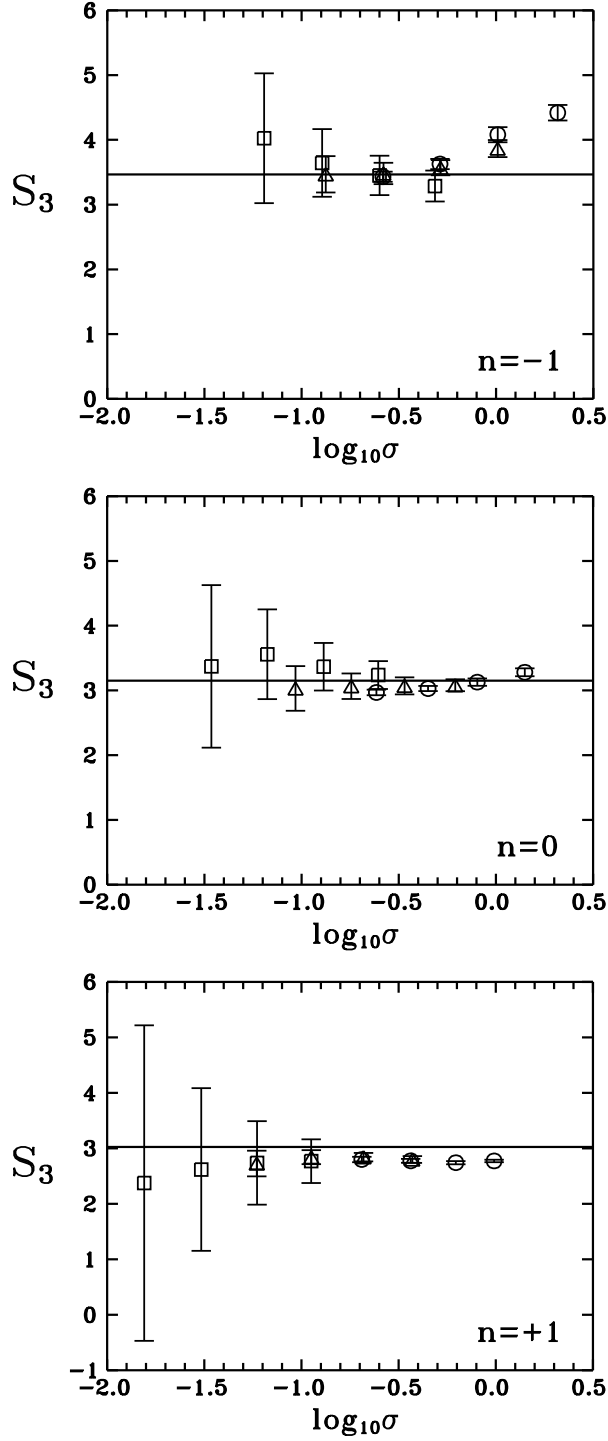

**Figure 1:** Results of the N–body simulations for different power spectra. Squares, triangles and circles represent the values of skewness of the density field with Gaussian smoothing lengths of $L/12.5$, $L/25$, and $L/50$, respectively, where $L$ is the size of the simulation cube. Attached to them are 1–$\sigma$ statistical error bars computed by taking the dispersion of values from the $N_{sim} = 8$ simulations and dividing by $\sqrt{N_{sim} - 1} = \sqrt{7}$. The solid lines show the results of perturbative calculations.

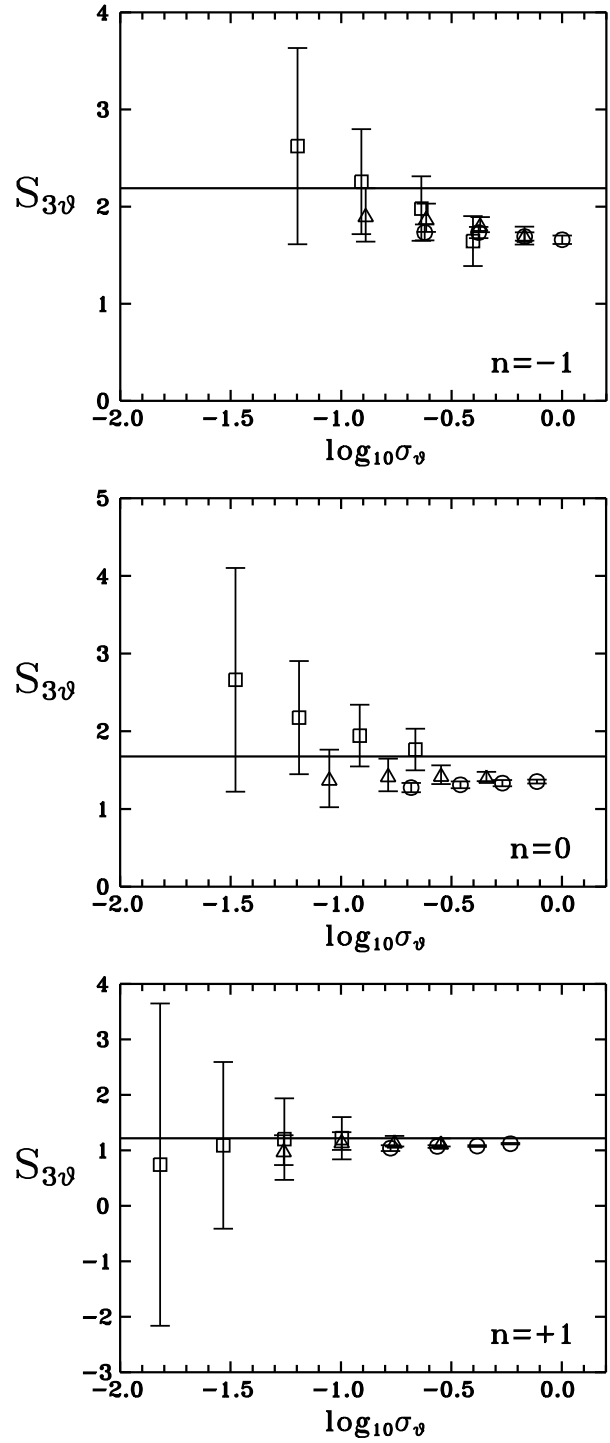

**Figure 2:** The same as Figure 1, for the absolute value of skewness of the velocity divergence field.

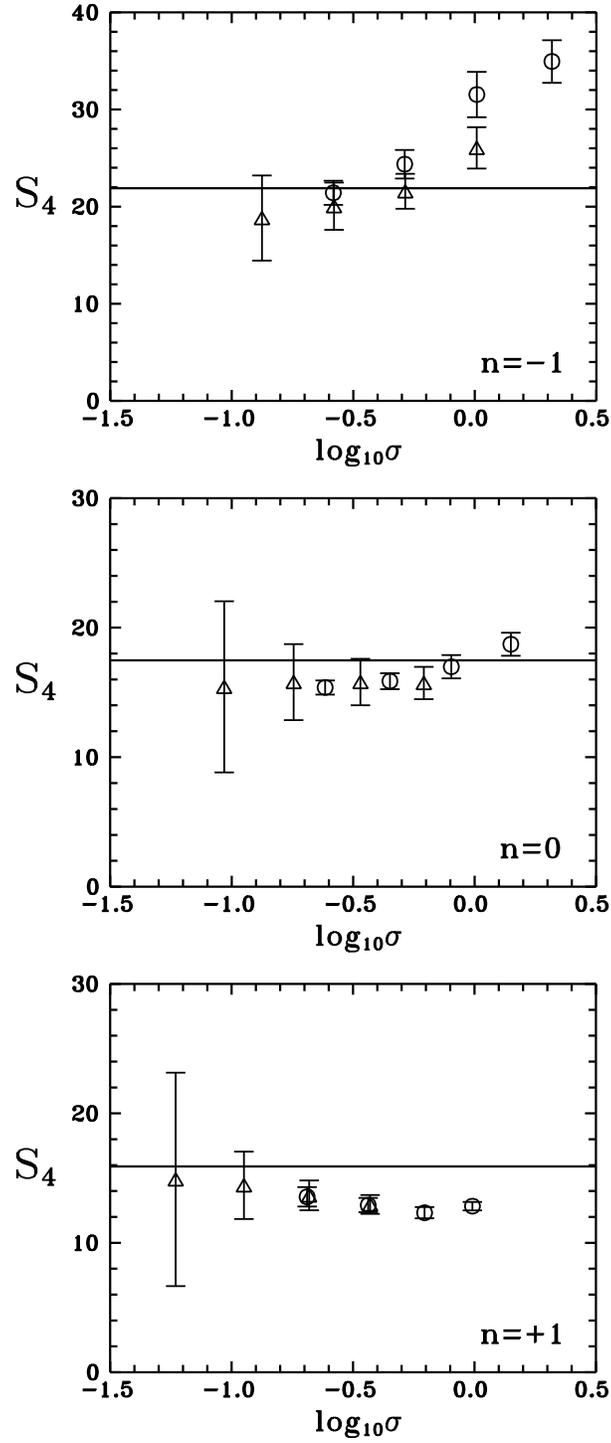

**Figure 3:** The same as Figure 1, for the kurtosis of the density field, except that only the results for the Gaussian smoothing lengths of $L/25$ and $L/50$ are shown.

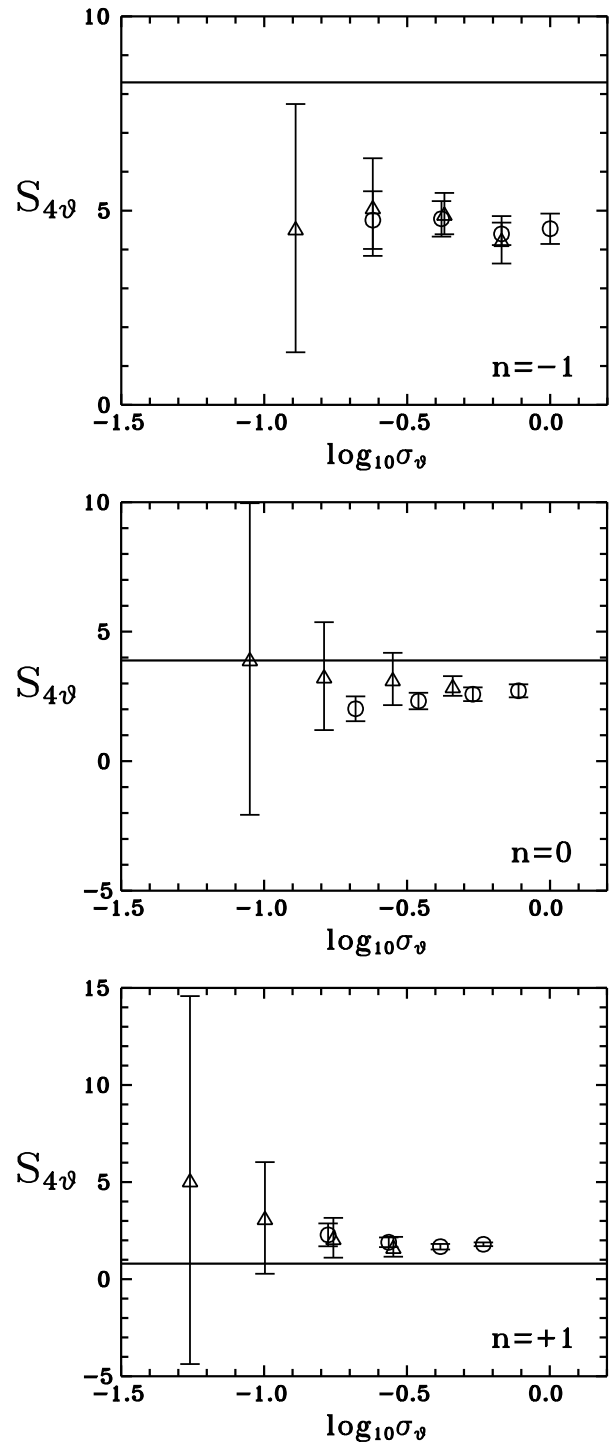

**Figure 4:** The same as Figure 3, for the kurtosis of the velocity divergence field.

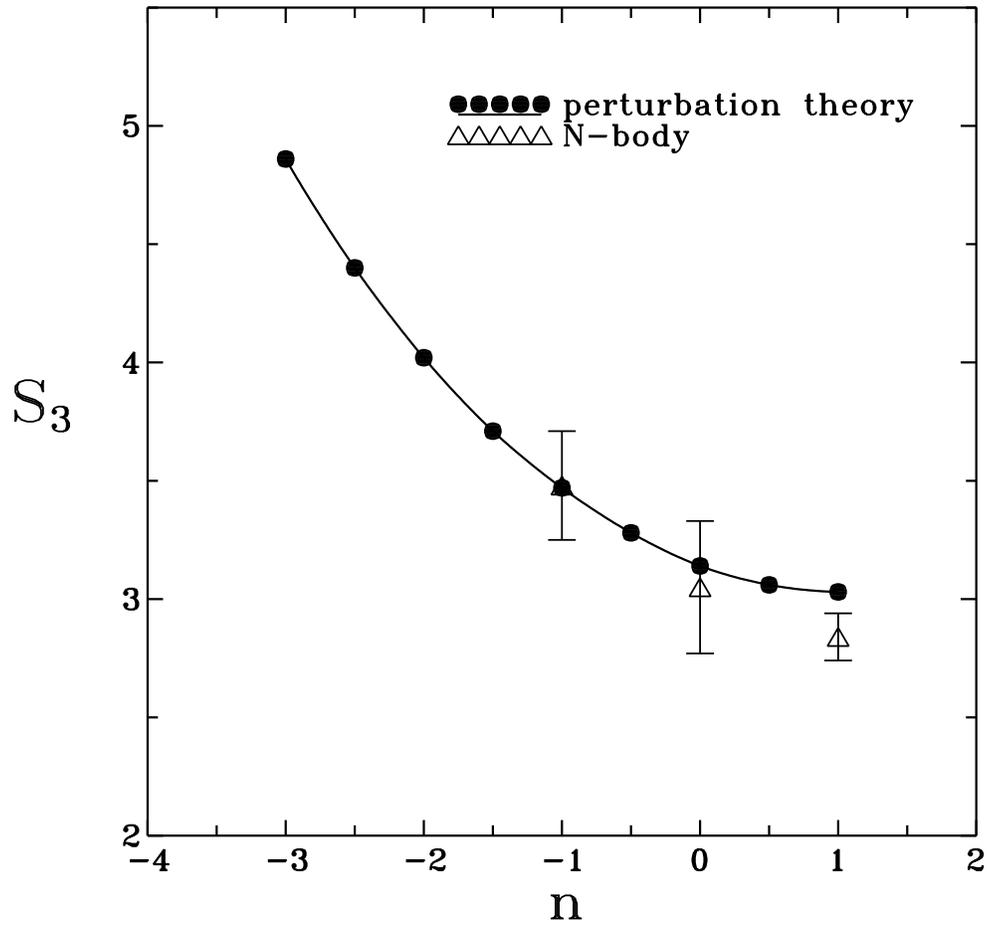

**Figure 5:** The dependence of the skewness of the density field on the spectral index $n$.

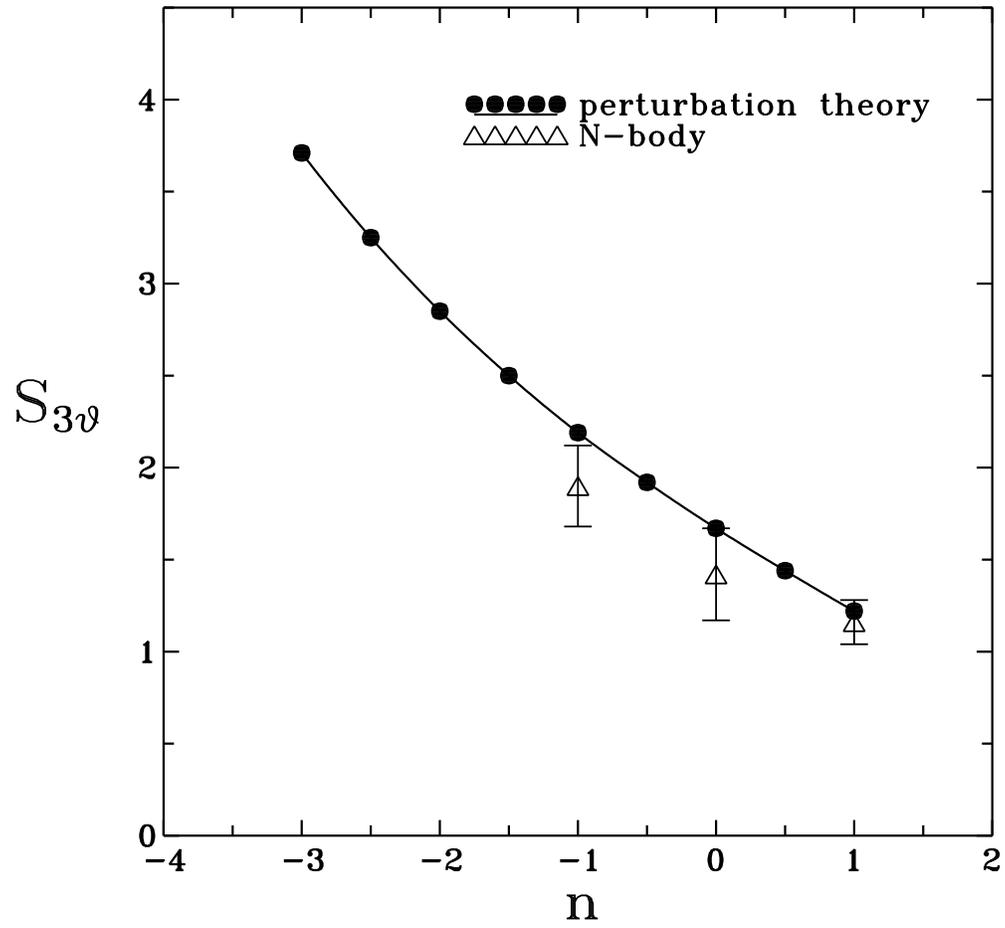

**Figure 6:** The dependence of the absolute value of skewness of the velocity divergence field on the spectral index $n$ for $\Omega = 1$.

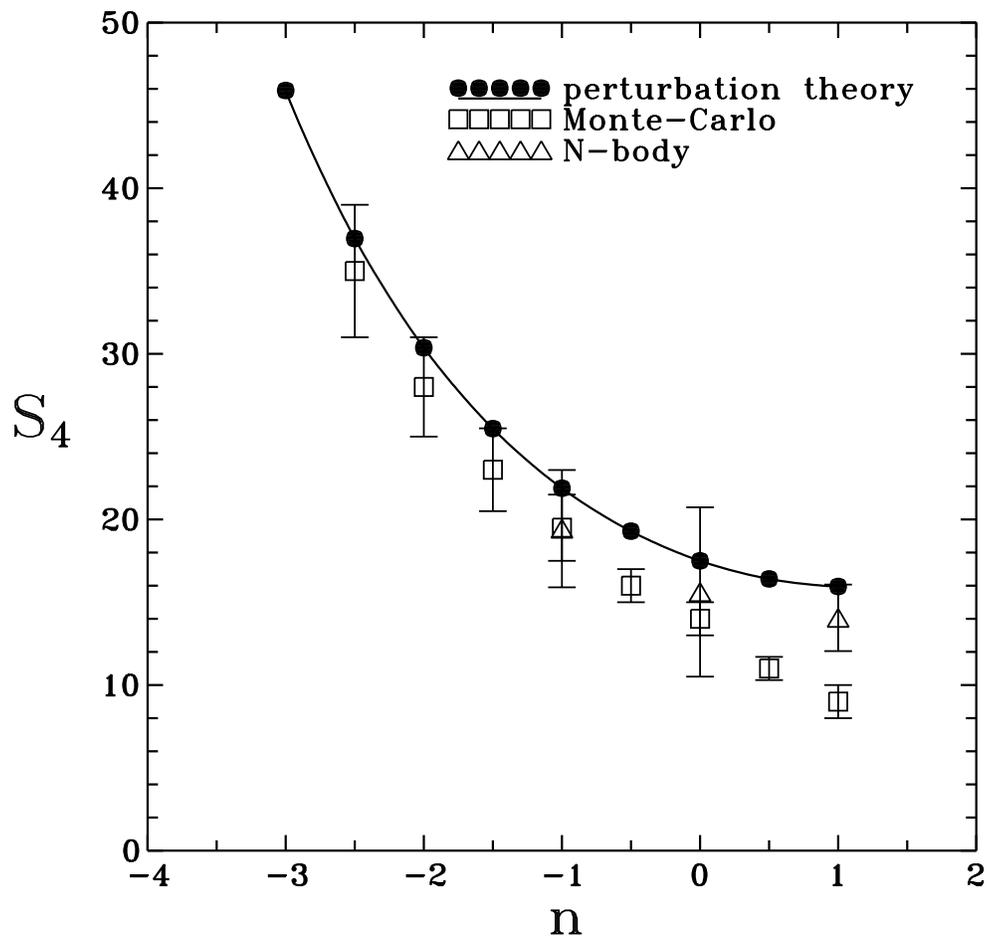

Figure 7: The dependence of the kurtosis of the density field on the spectral index $n$.

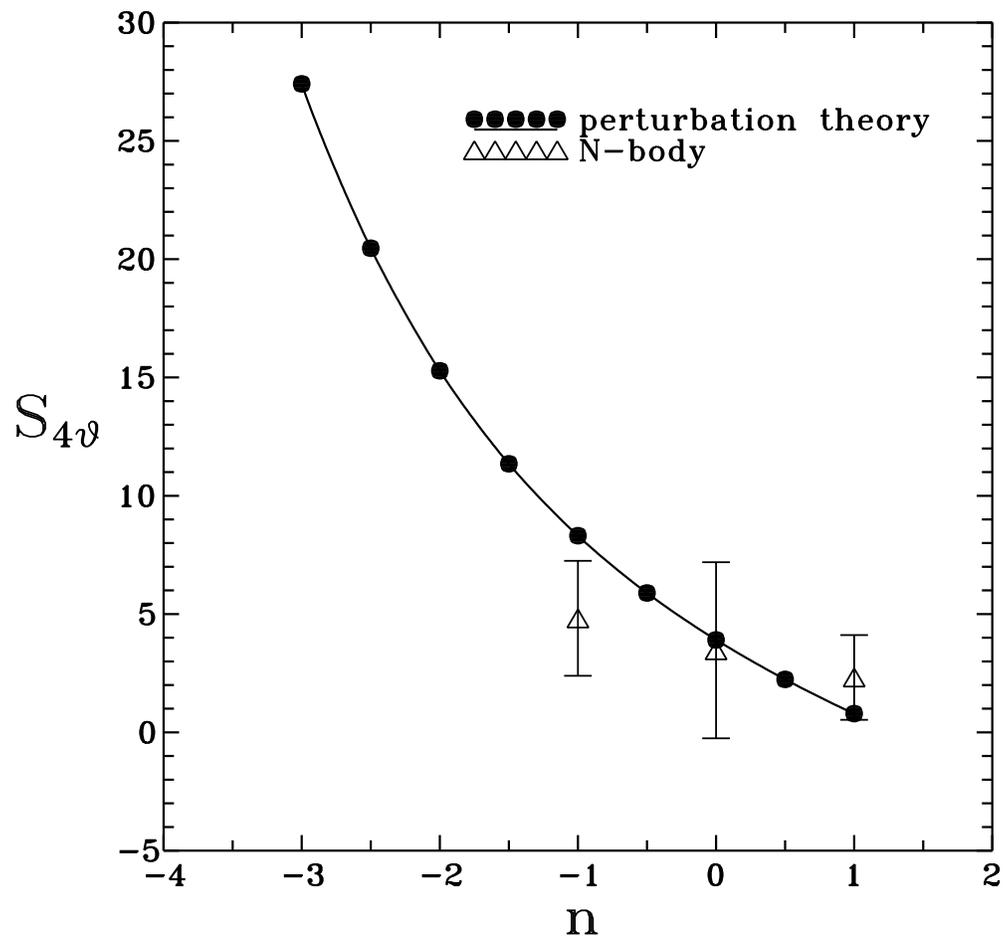

**Figure 8:** The dependence of the kurtosis of the velocity divergence field on the spectral index $n$ for $\Omega = 1$.